A Holistic, Non-algorithmic View of Cultural Evolution:
Commentary on Review Article by Prof. Liane Gabora for Physics of Life Reviews

Stuart Kauffman
Email: stukauffman@gmail.com

This is an outstanding paper. There is surely some truth to the notion that culture evolves, but the Darwinian view of culture is trivial. Consider the following example: the economic evolution from the Turing machine, to the mainframe computer, to personal computer, to word processing, file sharing, the Web, and selling things on the web. In what sense is selling on the web a mutant variant of the Turing machine? As Gabora points out, ideas and artifacts get put to new uses and combined with one another in new ways for new functionalities, and this is what underlies technological, cultural and political evolution. None of this is captured or even approachable by way of a Darwinian theory of culture.

Gabora does two things in this paper. First, she levels a reasoned and devastating attack on the adequacy of a Darwinian theory of cultural evolution, showing that cultural evolution violates virtually all prerequisites to be encompassed by Darwin's standard theory. Second, she advances the central concept that it is whole world views that evolve. A world view emerges when the capacity of memories to evoke one another surpasses a phase transition yielding a richly interconnected conceptual web, a world view. She proposes that cultural evolves not through a Darwinian process such as meme theory, but through communal exchange of facets of world views. Each section of her argument is completely convincing.

I would add only two suggestions that Gabora might consider. First, the philosophy of science since Popper was devastated by Quine years ago in a way that is entirely consistent with her "world view" as a whole picture. Scientists (of which I am one) tend to ignore philosophy, sometimes at their peril. Popper famously tried to solve Hume's problem of induction by claiming that scientists do NOT carry out induction. Rather, scientists put forth hypotheses, and what makes hypotheses scientific is their capacity to be falsified. Here Popper is in part driven by the fact that the universal claim, "All swans are white", can be disconfirmed by the observation of one black swan.

But Quine (1953) showed Popper to be wrong; given a theory and an experiment that appears to disconfirm the theory, we must give up SOME component of the theory, some hypothesis, but it is our CHOICE which one to give up. We tend to abandon the hypothesis that least disturbs the netted web of our hypotheses about the world, i.e., our (in Gabora's terms) world view. We tend to protect its overall structure as long as possible. Quine's case is compelling, and remains unknown to two generations of scientists who routinely ask of a research proposal, "How can the hypothesis be falsified?" In fact we can hold a given hypothesis safe against disproof, i.e., treat it as *a prioi* true, although at the price of a strange idea, e.g., quantum physics. This ADDs to Gabora's paper. It demonstrates a tenet of Gabora's theory that is ignored by Darwinian theorists of culture: that elements of culture (e.g., scientific hypotheses) do not evolve as separate entities but as facets of an integrated whole (a world view). This is revealed in many if not all aspects of human life including how we do science.

My next point is inspired by the argument presented in a recent paper in which my colleagues and I claim to have broken the reductive dream since Newton of a law that entails all that becomes in the universe (Longo, Montevil, & Kaufman, 2012). The basic argument can be explained by way of an example. If you attempt to list all the possible uses of a screwdriver your list might include: screw in a screw, open a can of paint, stab an attacker, and so forth. Upon reflection, the number of uses of a screwdriver is infinite. Moreover, unlike the orderable integers, the uses of a screwdriver are un-orderable. These premises imply that NO EFFEECTIVE PROCEDURE OR ALGORITHM can calculate or list all the uses of a screwdriver nor, in general, find a new use of a screwdriver. This is the famous FRAME problem of Computer Science, unsolved since Turing.

All that has to happen in an evolving bacterium in some new environment is that (1) some "molecular screwdriver" of sorts finds a use that enhances the fitness of that organism, and (2) there be heritable variation for that use. Then it may evolve that new use. How this "arrival of the Fitter" occurs was never solved by Darwin, nor by NeoDarwinism; indeed it cannot be solved because the new use is not pre-state-able. The new use changes the very phase space of possible evolutionary trajectories. But since the time of Newton we can only write laws of motion and integrate them if we can pre-state the phase space and the boundary conditions that define that phase space. We can do neither for the evolution of the biosphere, nor for the evolution of culture. Thus, the evolution of the biosphere and a fortiori, the evolution of the economy and culture are entailed by no physical laws.

These points are relevant to Gabora's Conceptual Network computer model of artifact evolution. Unlike previous such computer programs that use a finite string of symbols to represent the features of artifacts, this program allows the user to propose new attributes, and to weight the importance of hierarchical clusters of related attributes with respect to a particular goal or function using the notion of 'perspective'. Thus her approach allows for consideration of USES of artifacts that are not algorithmically list-able, nor even mathematizable in general (e.g., the many possible uses of a screwdriver). Looked at superficially, this may appear to be stacking the odds to generate the cultural lineage that one favors. But so long as the rationale for the included attributes and perspectives is stated beforehand, careful consideration of this move reveals it to be indispensible due to frame problem. Economic, technological, and cultural innovations are very often not recombinations of old objects with established uses, rather they are the non-deductive, non-algorithmic discovery of new uses, e.g., tie the screwdriver to a stick to spear fish and rent to natives for 5% of the catch. I suggest that Gabora's implicit acceptance that the phase space is not pre-state-able by allowing humans to introduce context-relevant attributes and perspectives is of particular importance, and should be highlighted.

The notion of a pre-stated opportunity space is central to Competitive General Equilibrium and current theories of price formation. The deep problems with this notion identified in Gabora's paper and discussed further here suggest that we need to alter how we think of economic "search" as if it were on a space of pre-stated possibilities, and think instead in terms of how new possibilities emerge through the interaction of self-organizing conceptual webs as described toward the end of Gabora's paper.

My goal here is to point to what I see as some of the most exciting implications of Gabora's excellent ideas. In short, it is an outstanding paper.

References

Gabora, L. (2013). An evolutionary framework for culture: Selectionism versus communal exchange. Physics of Life Reviews, 10(2), 117-145.

Longo, G., Montevil, M., & Kaufman, S., 2012. No entailing laws, but enablement in the evolution of the biosphere. In: *Proceedings of the Fourteenth International Conference on Genetic and Evolutionary Computation Conference*, pp. 1379-1392.

Quine, W. V. O. (1953) Two dogmas of empiricism. In: *From a logical point of view*. Cambridge MA: Harvard University Press.